\documentclass[twocolumn,epjc3]{svjour3}          

\RequirePackage[T1]{fontenc}

\smartqed  

\RequirePackage{graphicx}
\RequirePackage{flushend}
\RequirePackage[numbers,sort&compress]{natbib}
\RequirePackage[colorlinks,citecolor=blue,urlcolor=blue,linkcolor=blue]{hyperref}

 \usepackage[usenames,dvipsnames]{pstricks}
 \usepackage{epsfig}
 \usepackage{pst-grad} 
 \usepackage{pst-plot} 
\usepackage{cuted}

\usepackage{latexsym ,amsmath,amsfonts,amssymb,amstext,mathrsfs} 
\usepackage{enumerate}

\usepackage{slashed}

\usepackage{simplewick}
\usepackage{multirow}

\begin{document}

\title{Standard Model and New Physics contributions to $K_L$ and $K_S$ into four leptons}


\author{Giancarlo D'Ambrosio\thanksref{e1,addr1}
        \and
        David Greynat\thanksref{e2,addr1,addr2}
        \and 
        Gr\'egory Vulvert\thanksref{e3,addr3}
}

\thankstext{e1}{e-mail: gdambros@na.infn.it}
\thankstext{e2}{e-mail: greynat@na.infn.it}
\thankstext{e3}{e-mail: vulvert@ific.uv.es}

\institute{INFN-Sezione di Napoli, Via Cintia, 80126 Napoli, Italia\label{addr1}
          \and
          Dipartimento di Scienze Fisiche, Universit\'a di Napoli Federico II, Via Cintia, 80126 Napoli, Italia \label{addr2}
          \and
          Departament de F\'{i}sica Te\`{o}rica, IFIC, Universitat de Val\`{e}ncia - CSIC, 
Apt. Correus 22085, E-46071 Val\`{e}ncia, Spain \label{addr3}
}

\date{IFIC/13-66}

\maketitle

\abstract{
We study the $K_L$ and $K_S$  decays into four leptons ($e\bar{e}e\bar{e}, \, \mu\bar{\mu}\mu\bar{\mu},\, e\bar{e}\mu\bar{\mu}$) where we use a form factor motivated by vector meson dominance, and show the dependence of the branching ratios and spectra  from the slopes. A precise determination of short distance contribution to $K_L\to \mu\mu$ is affected by our ignorance on the sign of the amplitude $\mathcal{A}(K_L\to \gamma\gamma)$ but we show a possibility to measure the sign of this  amplitude by studying $K_L$ and $K_S$  decays in four leptons. We also investigate the effect of  New Physics contributions for  these decays }
%

\section{Introduction}

The recent LHCb measurement on $K_S \rightarrow\mu\overline{\mu}$~\cite{Aaij:2012rt} 
is getting closer to the Standard Model (SM) prediction
\begin{align}
& \text{Br }(K_S \rightarrow\mu\overline{\mu})\big \vert_\text{LHCb}  < 9\times 10^{-9} \  {\rm at} \  90    \  \% \   {\rm CL}  \\
& \text{Br }(K_S \rightarrow\mu\overline{\mu})\big \vert_\text{SM}  =(5.0\pm 1.5) \times 10^{-12},
\label{eq:KSmumuexp}\end{align}
and this has motivated our interest in studying other feasible decays at LHC~\cite{Bediaga:2012py}  or other facilities: decays of $K_{L,S}$ into two Dalitz pairs ($K_{L,S} \rightarrow\mu\overline{\mu}\mu\overline{\mu},$ $e\overline{e}  \mu\overline{\mu}, \ e\bar{e}e\bar{e}$).
These decays have received attention before. Compared to the previous literature~\cite{Miyazaki,Ecker:1991ru,cappiello,Goity,Birkfellner,Cirigliano:2011ny}, in this paper we have introduced a form factor, motivated by vector meson dominance and a good behaviour at short distance~\cite{DAIP}, which is particularly important  for $K_{L,S} \rightarrow\mu\overline{\mu}\mu\overline{\mu}$, 
the one more easily detectable at LHCb. We study the dependence of the spectra and the branching ratio from the linear and quadratic slopes of the form factor. 

We  show also that the measurement of the  time interference of $\mathcal{A}(K_{L} \to \ell \bar{\ell} \ell\bar{\ell}) $ with $\mathcal{A}(K_{S} \to \ell \bar{\ell} \ell \bar{\ell})$ would allow the determination of the sign  of $\mathcal{A}(K_{L} \to \gamma \gamma) $, this observable indeed depends linearly from $\mathcal{A}(K_{L} \to \gamma \gamma) $. This experimental  determination is very welcome since would allow CKM stringent tests~\cite{IU}.

We also discuss    two possible New Physics (NP) models that  can be studied by measuring measurements of $\mathcal{A}(K_{L,S} \to \ell \bar{\ell} \ell \bar{\ell}) $:
\begin{enumerate} [\itshape(i)]
	\item a direct NP coupling for $K_L \gamma \gamma$.
	\item a Bremsstrahlung part from $K_{L,S} \rightarrow\mu\overline{\mu}$.
\end{enumerate}

We discuss in order:  the chiral perturbation theory (ChPT) and vector meson dominance (VMD) description of $K_{L,S}$ decays in section 2 and, in  section 3, the results associated (including the kinematics). The different possibilities of interferences are discussed in section 4 including the Bremsstrahlung contributions and the CP-violation in the $K_S$ decays. The appendix contains some detailed expressions for the amplitudes. 



\section{Chiral perturbation theory  description of $K_{L,S}\to \bar{\ell} \ell  \bar{\ell} \ell$}

\subsection{$K_{L}\to \bar{\ell}  \ell  \bar{\ell} \ell$ }
$K_{L} \to  \mu^+\mu^-$ decay receives large   long distance (LD) contributions and small  short  distance (SD) contributions:
to disentangle the small  but interesting  short   distance contribution 
an accurate description of the long distance contribution $K_{L} \to \gamma (q_{1}) \gamma (q_{2})\to \mu^+\mu^-$
is required. To this purpose the authors of ref.  \cite{DAIP} introduce  a   form factor $\mathrm{F}_L(q_1^2, q_2^2) $ motivated by
the assumption that VMD plays a crucial role in the matching between short and long distances
\begin{multline}
\mathrm{F}_L(q_1^2, q_2^2) \doteq \mathrm{F}_L(0,0) \left[1 + \alpha_L  \left(\frac{q_1^2}{q_1^2-M_V^2} + \frac{q_2^2}{q_2^2-M_V^2} \right) \right. \\
\left.+ \beta _L \frac{q_1^2 q_2^2}{(q_1^2-M_V^2)(q_2^2-M_V^2)} \right]\label{eq:Ftz},
\end{multline}
$\mathrm{F}_L(0,0)$ is a constant fixed by the experimental width $\Gamma(K_{L} \to \gamma  \gamma )$. The duality properties of this form factor are implemented by determining possibly  $\alpha _L$ and $\beta _L$ in the  low energy expansion from experiments and imposing a phenomenological matching with the QCD short distance result \cite{DAIP}. We match the SD non-zero value  with the  form factor at short distance  
\begin{equation}
1+2\alpha_L+\beta_L=0.3. \label{eq:sumrule}
\end{equation}
As shown in~\cite{DAIP},  experiments, mainly from $K_{L} \to \gamma   \gamma^*$ decay,
  fix  the value of $\alpha_L=-1.69 \pm 0.08 $ \cite{PDG}, while  the   experimental determination of   $\beta  _L$ from $K_L\to \bar{\ell}  \ell \bar{\ell} \ell$ would allow a test of saturation with one resonance ($\rho$)
 of the sum rule in eq. (\ref{eq:sumrule}). 
Since this experimental determination is still missing  either we rely on $\beta  _L$ from eq.(\ref{eq:sumrule}) 
 or, as we will do, we plot  ${\rm Br} (K_L\to \bar{\ell} \ell \bar{\ell} \ell)$ as function of $\beta  _L$ in  figure \ref{fig:KL4lnorm}.
 
 The value of $\mathrm{F}_L(0,0)$ is   fixed through the amplitude of $K_L  \to \gamma^*(q_1, \epsilon_1) \, \gamma^*(q_2, \epsilon_2)$ 
 \begin{equation}
\mathcal{A}\left(K_L  \to \gamma^* \, \gamma^* \right) = i \varepsilon_{\mu \nu \rho \sigma } q_1^\rho q_2^\sigma \, \epsilon_1^\mu \, \epsilon_2^\nu \, \mathrm{F}_L(q_1^2, q_2^2),
\end{equation}
with the effective lagrangian 
\begin{equation}
\mathcal{L}_{\mathrm{eff}} = -\frac{\mathrm{F}_L(0,0)}{8} \, \varepsilon_{\mu \nu \rho \sigma } \, K_2 F^{\mu \nu} F^{\rho \sigma}+ \text{h.o.t.},
\end{equation}
we can directly connect $\mathrm{F}_L(0,0) $ to the branching ratio 
\begin{align}
\Gamma (K_L \to \gamma \gamma) &= \frac{M_K^3 |\mathrm{F}_L(0,0)|^2}{64 \pi}\nonumber
\\
&=(7.16 \pm 0.05) \times 10^{-21} \text{ GeV},
\end{align}
and therefore (for the numerical evaluation, we will use the central value only) ,
\begin{align}\label{eq:FFKL}
|\mathrm{F}_L(0,0)| &= \left[ \frac{64 \pi \, \Gamma(K_L \to \gamma \gamma)}{M_K^3}\right]^{1/2} \nonumber \\
 &=( 5.61 \pm 0.06) \times 10^{-9}\text{ GeV}^{-1}.
\end{align}

\subsection{$K_{S}\to \bar{\ell}  \ell \bar{\ell}\ell$}
The first non-trivial  ChPT contribution to $K_S\to \gamma  ^* \gamma ^*$ appears at ${\cal O }(p^4)$: no counterterms are allowed by chiral symmetry at this order implying that chiral loops are finite \cite{Ecker:1991ru}.  In this paper  we want to account for two important ${\cal O}(p^6)$ effects: 
\begin{enumerate} [\itshape(i)]
\item We need to add  local ${\cal O}(p^6)$  contributions to  the ${\cal O }(p^4)$ $K_S\to \gamma    \gamma$  to match exactly   the experimental value.
\item Potentially important vector meson dominance contribution ${\cal O}(p^6)$ to $K_S\to \gamma  ^* \gamma ^*$ generated by  the ${\cal O }(p^4)$ electromagnetic form factor of the pion.
\end{enumerate}
 We discuss the strategy to account for these effects. Writing the amplitude of $K_S  \to \gamma^*(q_1, \epsilon_1) \, \gamma^*(q_2, \epsilon_2)$
\begin{multline}
\mathcal{A}(K_S  \to \gamma^*\gamma^*) \\
=  i \left[ (q_1 \cdot q_2) g^{\mu \nu} - q_2^\mu q_1^\nu \right] \, \epsilon_{1 \mu} \, \epsilon_{2 \nu} \, \mathrm{F}_S(q_1^2, q_2^2),
\end{multline}
we can obtain the PDG experimental value $\mathrm{F_S}(0,0)$ \cite{PDG},
\begin{align}
 \Gamma (K_S \to \gamma \gamma) &= \frac{M_K^3 |\mathrm{F}_S(0,0)|^2}{64 \pi} \nonumber \\
 & =(1.93 \pm 0.12)\times 10^{-20} 
 \end{align}
adding an ${\cal O}(p^6)$ local term  to   the ${\cal O }(p^4)$ $K_S\to \gamma    \gamma  $ chiral loop  \cite{DEG} as done  
 in ref. \cite{Buchalla:2003sj}, then (here too, we will use only the central value for the numerical evaluation)
\begin{align}
|\mathrm{F}_S(0,0)| &= \left[ \frac{64 \pi \, \Gamma(K_S \to \gamma \gamma)}{M_K^3}\right]^{1/2} \nonumber\\
&= (3.38 \pm 0.03 )\times10^{-9} \text{ GeV}^{-1}.\label{eq:FFKS}
\end{align}

We want also to add the potentially important vector meson dominance contribution  ${\cal O}(p^6)$ to $K_S\to \gamma  ^* \gamma ^*$: this is generated by  the ${\cal O }(p^4)$ electromagnetic form factor of the pion; this problem was already studied  to evaluate the potentially important ${\cal O}(p^6)$ VMD  contribution to  $K_L\to \pi^0  \gamma  ^* \gamma ^*$ (\cite{Buchalla:2003sj} and references therein). 
The leading chiral contribution to  $K_L\to \pi^0 \gamma   \gamma $ appears  at ${\cal O }(p^4)$: no counterterms are allowed by chiral symmetry at this order implying that chiral loops are finite.  Large VMD  and unitarity ${\cal O}(p^6)$  corrections  to  $K_L\to \pi^0 \gamma   \gamma $, as required by phenomenology have been investigated \cite{cappiello}. In ref.~\cite{Buchalla:2003sj} the effects of the pion  electromagnetic form factor to the pion loop amplitude 
$\mathcal{A}(K_L\to \pi^0 \gamma  ^*(q_1) \gamma ^* (q_2)) $ have been studied: they suggest to approximate this amplitude as the product of the  amplitude with the photons on shell  multiplied  a form factor like the one in eq.~~(\ref{eq:Ftz}).
Very similarly to ref.~\cite{Buchalla:2003sj} (and references therein)
to include this  VMD contribution we  suggest to approximate the full amplitude as
\begin{multline}
\mathrm{F}_S(q_1^2, q_2^2) \doteq \mathrm{F}_S(0,0) \left[1 + \alpha_S  \left(\frac{q_1^2}{q_1^2-M_V^2} + \frac{q_2^2}{q_2^2-M_V^2} \right) \right. \\
\left.+ \beta _S \frac{q_1^2 q_2^2}{(q_1^2-M_V^2)(q_2^2-M_V^2)} \right].
\end{multline}
where $\mathrm{F}_S(0,0)$ is the  ${\cal O}(p^4)$ $K_S\to \gamma   \gamma$     chiral loop loop amplitude, with on-shell photons,   plus a local ${\cal O}(p^6)$ as discussed in connection with eq. (\ref{eq:FFKS}).
Differently  from $K_L\to \gamma  ^* \gamma ^*$, sum rule in eq. (\ref{eq:sumrule}) due to the vanishing SD contribution
 the limit $q_{1,2}^2 \gg M_V^2$ imposes the constraint~\cite{DAIP}:
\begin{equation}
1+2\alpha_S+\beta_S=0,\label{eq:SDC2}
\end{equation}
reducing the number of unknown parameters to one. 

In principle the off-shell photon behavior of  ${\cal O }(p^4)$  $\mathcal{A}(K_S\to   \gamma  ^*(q_1) \gamma ^* (q_2)) $  from ref. \cite{Ecker:1991ru}  could affect $\alpha _S $,   $\beta _S $ or add other gauge invariant structures but we have checked that these effects are negligible\footnote{Numerically we have found that these  effects generate  $\alpha _S $ and  $\beta _S $ at  $\mathcal{O}(0.1)$, other effects are substantially smaller. Also we have checked  that our parametrization of the off-shell photon behavior of  $\mathcal{O }(p^4)$  of  $\mathcal{A}(K_S\to   \gamma ^*\gamma ^*) $  from ref.~\cite{Ecker:1991ru}  in terms of $\alpha _S $ and    $\beta _S $  reproduce well  ${\rm Br}(K_{S } \to 4\ell )$ as described in  ref.~\cite{Birkfellner}.} to potentially large effects from VMD.

\section{Kinematics and results}\label{KS}

\subsection{$K_L \to \ell_1 \bar{\ell}_1 \ell_2 \bar{\ell}_2$}

\begin{figure*}
\begin{center}
\scalebox{0.7} 
{
\begin{pspicture}(0,-3.312)(19.325907,3.312)
\definecolor{color28g}{rgb}{0.9176470588235294,0.9490196078431372,0.9215686274509803}
\definecolor{color28f}{rgb}{0.7215686274509804,0.9764705882352941,0.796078431372549}
\definecolor{color28}{rgb}{0.10980392156862745,0.9450980392156862,0.09411764705882353}
\definecolor{color29}{rgb}{0.4,0.8941176470588236,0.17254901960784313}
\definecolor{color35g}{rgb}{0.8470588235294118,0.8627450980392157,0.9686274509803922}
\definecolor{color35f}{rgb}{0.9333333333333333,0.9333333333333333,0.9333333333333333}
\definecolor{color35}{rgb}{0.07058823529411765,0.12941176470588237,0.8392156862745098}
\definecolor{color36}{rgb}{0.4235294117647059,0.9137254901960784,0.17647058823529413}
\definecolor{color39}{rgb}{0.8627450980392157,0.10980392156862745,0.1568627450980392}
\definecolor{color40}{rgb}{0.3803921568627451,0.22745098039215686,0.8392156862745098}
\definecolor{color46}{rgb}{0.8,0.16862745098039217,0.16862745098039217}
\definecolor{color53}{rgb}{0.807843137254902,0.0784313725490196,0.0784313725490196}
\definecolor{color54}{rgb}{0.11764705882352941,0.03529411764705882,0.6274509803921569}
\definecolor{color60}{rgb}{0.2901960784313726,0.9098039215686274,0.13725490196078433}
\definecolor{color64}{rgb}{0.7019607843137254,0.054901960784313725,0.054901960784313725}
\pspolygon[linewidth=0.024,linecolor=color28,fillstyle=gradient,gradlines=2000,gradbegin=color28g,gradend=color28f,gradmidpoint=1.0](8.333906,2.74)(3.3539062,2.72)(0.3339062,-3.299406)(8.353907,-3.3)(9.753906,-0.36)
\psline[linewidth=0.024cm,linecolor=color29,linestyle=dashed,dash=0.16cm 0.16cm](4.373906,-3.2)(7.353906,2.78)
\psline[linewidth=0.03cm,arrowsize=0.113cm 2.04,arrowlength=1.45,arrowinset=0.38]{<->}(5.433906,0.96)(6.1939063,-1.4)
\usefont{T1}{ppl}{m}{n}
\rput(5.522969,1.23){$\mu^-$}
\usefont{T1}{ppl}{m}{n}
\rput(6.462969,-1.59){$\mu^+$}
\pspolygon[linewidth=0.024,linecolor=color35,fillstyle=gradient,gradlines=2000,gradbegin=color35g,gradend=color35f,gradmidpoint=1.0](8.053906,3.28)(15.873906,3.3)(19.313906,-3.22)(11.093906,-3.22)
\psline[linewidth=0.024,linecolor=color36,linestyle=dashed,dash=0.16cm 0.16cm](8.273906,2.752)(11.233906,2.7320526)(9.8,-0.208)(9.753906,-0.28)
\psline[linewidth=0.024cm,linecolor=color39,linestyle=dashed,dash=0.16cm 0.16cm](1.8339062,-0.24)(8.16,-0.288)
\psline[linewidth=0.024cm,linecolor=color40,linestyle=dashed,dash=0.16cm 0.16cm](12.073906,3.26)(15.193906,-3.2)
\psline[linewidth=0.03cm,arrowsize=0.113cm 2.04,arrowlength=1.45,arrowinset=0.38]{<->}(12.693906,-1.24)(15.173906,0.92)
\usefont{T1}{ppl}{m}{n}
\rput(15.432968,1.13){$e^-$}
\usefont{T1}{ppl}{m}{n}
\rput(12.4929695,-1.49){$e^+$}
\psline[linewidth=0.03cm,linecolor=color46,arrowsize=0.113cm 2.04,arrowlength=1.45,arrowinset=0.38]{->}(9.713906,-0.32)(8.193906,-0.32)
\usefont{T1}{ppl}{m}{n}
\rput(8.902969,-0.65){$K_{L,S}$}
\usefont{T1}{ppl}{m}{n}
\rput(1.4560938,-2.99){$\mu-\mu$ plane}
\usefont{T1}{ppl}{m}{n}
\rput(18.196095,-2.97){$e-e$ plane}
\psline[linewidth=0.024cm,linecolor=color53,linestyle=dashed,dash=0.16cm 0.16cm](9.8,-0.308)(17.7,-0.248)
\psarc[linewidth=0.024,linecolor=color54,arrowsize=0.093cm 2.02,arrowlength=1.38,arrowinset=0.24]{->}(14.04,-0.288){0.46}{0.0}{60.0}
\usefont{T1}{ppl}{m}{n}
\rput(14.754531,0.0020000006){$\theta_2$}
\psarc[linewidth=0.024,linecolor=color60,arrowsize=0.093cm 2.0,arrowlength=1.38,arrowinset=0.24]{<-}(5.76,-0.228){0.52}{99.46232}{185.19443}
\psarc[linewidth=0.03,linecolor=color64,arrowsize=0.093cm 2.02,arrowlength=1.38,arrowinset=0.24]{->}(9.72,-0.188){0.58}{60.0}{120.0}
\usefont{T1}{ppl}{m}{n}
\rput(9.744532,0.662){$\phi$}
\usefont{T1}{ppl}{m}{n}
\rput(4.974531,0.022){$\theta_1$}
\end{pspicture} 
}
\caption{Kinematics variables for the decays of the $K_{L,S}$ into 2 Dalitz pairs.}
\label{fig:geometric}
\end{center}
\end{figure*}
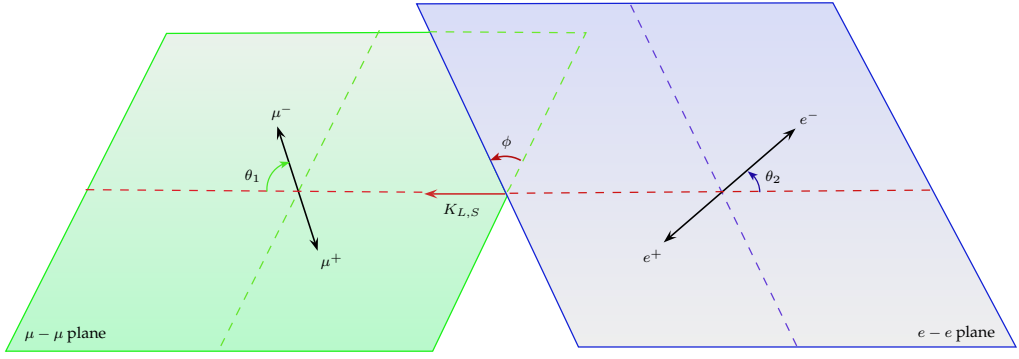

The cases that we calculated are $\ell_1 = \ell_2= e$, $\ell_1 = \ell_2= \mu$ and the composite case $\ell_1 = \mu$ and $\ell_2=e$ (see appendices for more detailed expressions). For each branching ratio, we have to use the phase space measure based on $5$ different variables completely determining the system. We choose two momenta and three angles. Thus we have to make a geometric treatment cf. fig.~\ref{fig:geometric}  and we will use the Cabibbo-Maksymovych approach~\cite{CabMak} 
\begin{equation}
\mathrm{d} \Phi_4 = \frac{\pi^2}{2^5 M_K^2} \, \sigma_1 \, \sigma_2 \, X \mathrm{d} q_1^2 \, \mathrm{d} q_2^2 \, \mathrm{d} (\cos \theta_1) \, \mathrm{d} \phi \,\mathrm{d} (\cos \theta_2),
\end{equation}
where
\begin{eqnarray}
X & = & \frac{1}{2} \lambda^{1/2}(M_K^2,q_1^2,q_2^2),\\
\sigma_i & = & \left( 1 - 4 \frac{m_i^2}{q_i^2} \right)^{1/2} , \, \, i= 1,2
\end{eqnarray}
and $\lambda$ is the well-known K\"all\`en function, 
\begin{equation}
\lambda(x,y,z) = (x-y-z)^2 - 4 yz.
\end{equation}
Here the integrations bounds are ($m$ here stands for the smallest mass between the leptons $\ell_i$):
\begin{equation}
\begin{cases}
4 m_\mu^2 \le q_1^2 \le (M_K-2m)^2, \\
4 m^2 \le q_2^2 \le (M_K-\sqrt{q_1^2})^2, \\
0 \le \theta_1, \theta_2 \le \pi, \\
0 \le \phi \le 2 \pi.
\end{cases}
\end{equation} 

Then, any differential decay width is given from the corresponding amplitude $\mathcal{M}$ by
\begin{equation}
\mathrm{d} \Gamma(K_L \to \ell_1 \bar{\ell}_1 \ell_2 \bar{\ell}_2)  =  \frac{(2 \pi)^{-8}}{2 M_K} |\mathcal{M}|^2 \mathrm{d} \Phi_4.
\end{equation}

\begin{table*}
\renewcommand{\arraystretch}{1.5}
\begin{center}
\caption{Results for the branching ratios of $K_L$ decays}
\begin{tabular*}{\textwidth}{@{\extracolsep{\fill}}lcccc@{}}
\cline{2-5}
\multicolumn{1}{c}{} & \multicolumn{2}{c}{$\alpha_L=\beta_L=0$} & $\alpha_L=-1.63$ and $\beta_L=0.3-1-2\alpha_L$ & Experiment \\
\cline{2-4}
\multicolumn{1}{c}{} & This work & Miyazaki \textit{et al.}~\cite{Miyazaki} & This Work &  PDG~\cite{PDG} \\ 
\hline
$K_L \to \mu \bar{\mu} \mu \bar{\mu}$ & $4.82 \times10^{-13}$ & $5.17 \times10^{-13}$ & $8.78\times10^{-13}$ & ----- \\ 
$K_L \to e \bar{e} e \bar{e}$ & $3.40 \times10^{-8}$ & $3.22 \times10^{-8}$ & $3.65 \times10^{-8}$ & $(3.56 \pm 0.21) \times10^{-8}$ \\ 
$K_L \to \mu \bar{\mu} e \bar{e}$ & $1.55 \times 10^{-9}$ & $7.77 \times 10^{-10}$ & $2.51 \times 10^{-9}$ & $(2.69 \pm 0.27) \times 10^{-9}$ \\
\hline
\end{tabular*}\label{tab:KLdecays}
\end{center}
\renewcommand{\arraystretch}{1}
\end{table*}

\begin{figure*}
\centering
\includegraphics[scale=0.85]{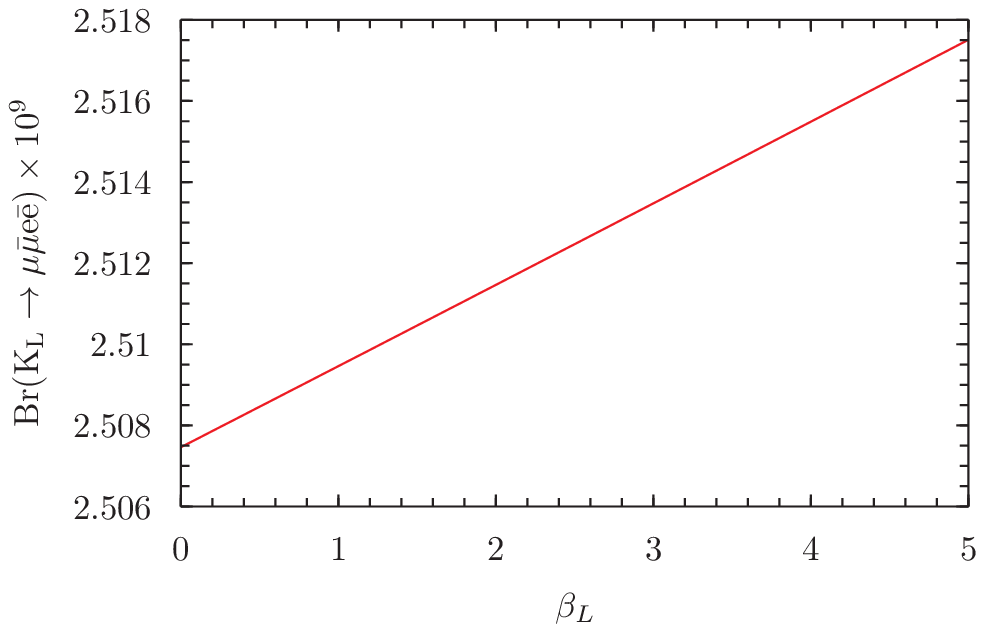}  \\ 
\centering
\vspace*{0.8cm}
\includegraphics[scale=0.85]{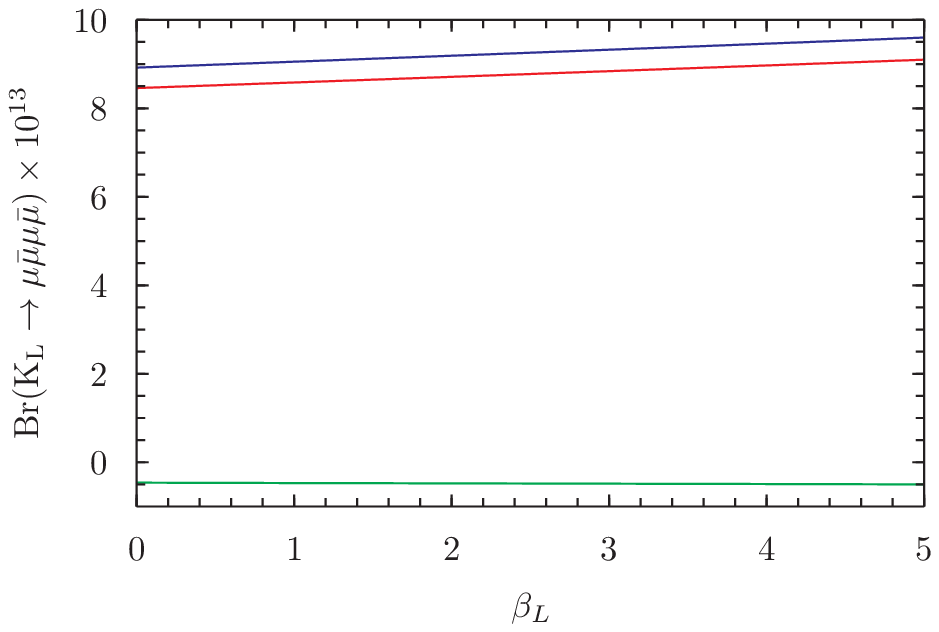}  \hspace*{0.8cm}\includegraphics[scale=0.85]{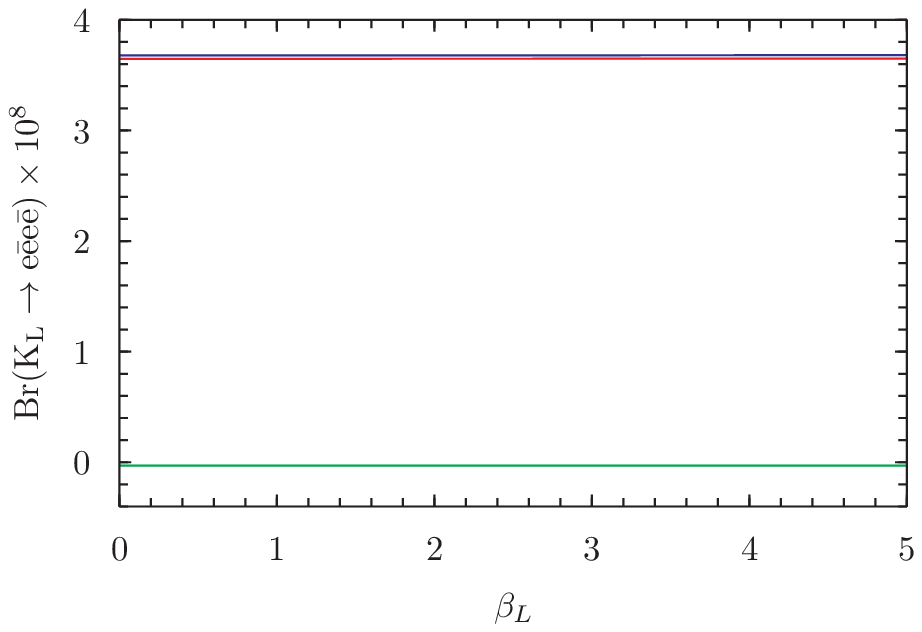}     
 \caption{Branching ratio of $K_L \to  \ell_1 \bar{\ell}_1 \ell_2 \bar{\ell}_2 $ vs. $\beta_L$. We fixed $\alpha_L=-1.63$~\cite{DAIP}. For the cases $\ell_1=\ell_2$, the red line is the total branching ratio, the blue one is the contribution of $|\mathcal{M}_A|^2$ and the green one is the contribution of the interference term $\mathcal{M}_A \mathcal{M}_B^*$.}
 \label{fig:KL4lnorm}
\end{figure*}

We give the results in table~\ref {tab:KLdecays} and the evolution of the various branching ratios according to $\beta_L$ is illustrated on fig.~\ref{fig:KL4lnorm}.

\subsection{$K_S \to \ell_1 \bar{\ell_1} \ell_2 \bar{\ell_2}$}

\begin{figure*}
\centering
\includegraphics[scale=0.85]{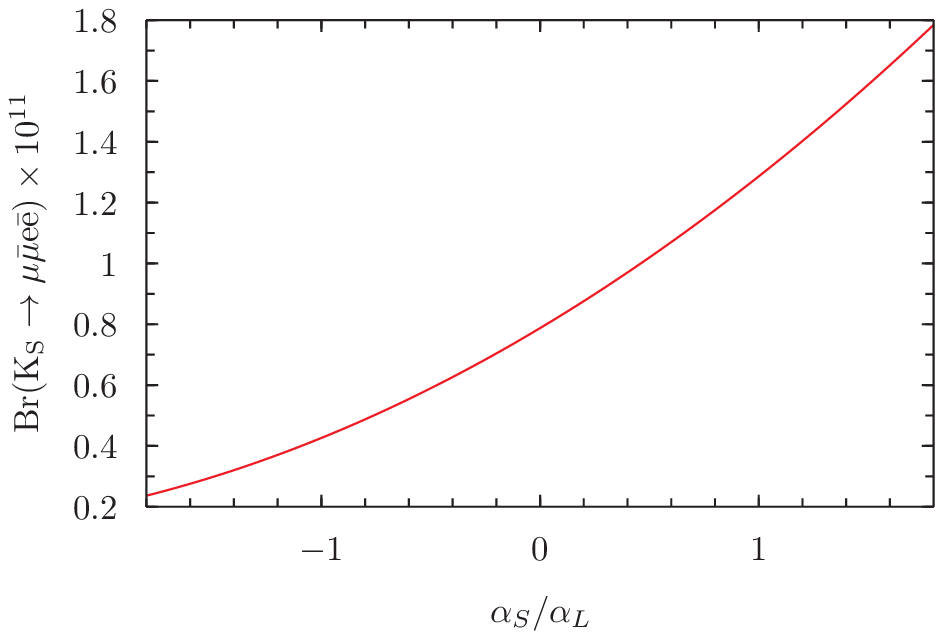}  \\ 
\centering
\vspace*{0.8cm}
\includegraphics[scale=0.85]{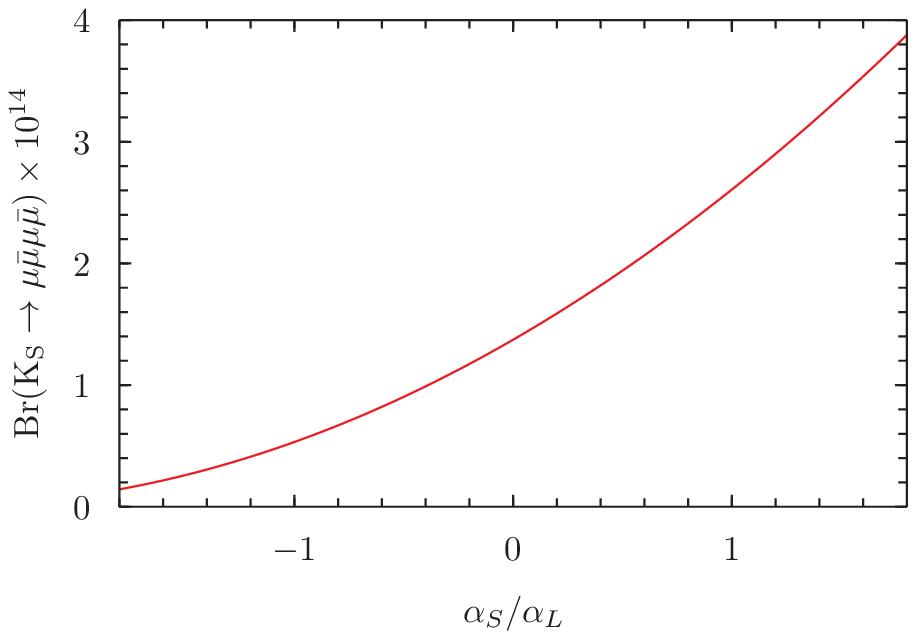} \hspace*{0.8cm}\includegraphics[scale=0.85]{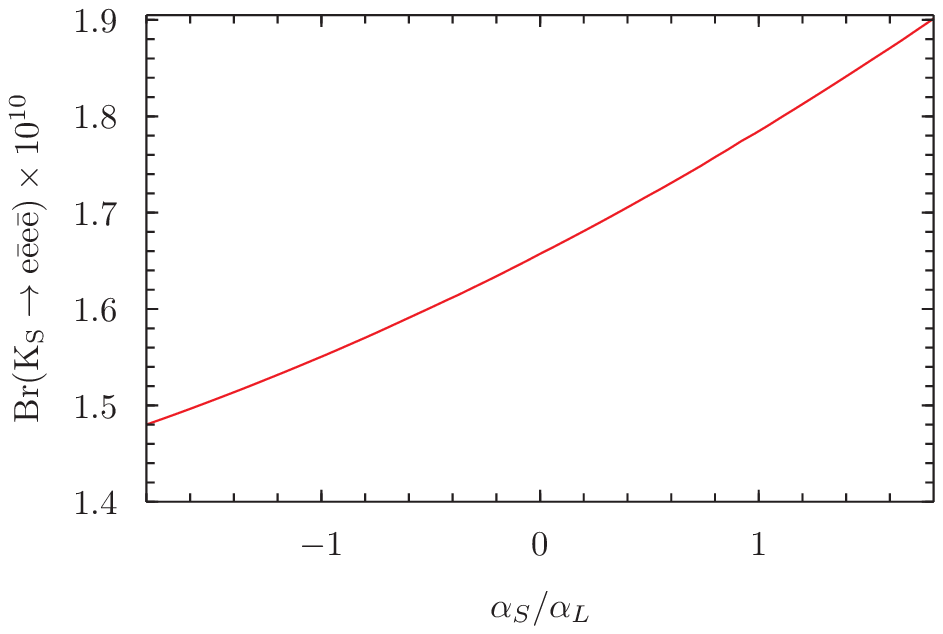}      
\caption{Branching ratio of $K_S \to  \ell_1 \bar{\ell_1} \ell_2 \bar{\ell}_2$ vs. $\alpha_S/\alpha_L$. We fixed $\alpha_L=-1.63$~\cite{DAIP} and $\beta_S=-1-2\alpha_S$ using~(\ref{eq:SDC2}).}
\label{fig:KS4lnorm}
\end{figure*}

In the same manner, we consider like for $K_L$ the cases $\ell_1 = \ell_2= e$, $\ell_1 = \ell_2= \mu$ and the composite case $\ell_1 = e$ and $\ell_2=\mu$ (see appendices). We present here our values for these decays in table~\ref{tab:KSdecays} and their values according to the variations of $\alpha_S/\alpha_L$ (cf. fig.~\ref{fig:KS4lnorm}). 

\begin{table*}[!]
\renewcommand{\arraystretch}{1.5}
\begin{center}
\caption{Results for the branching ratios of $K_S$ decays. Notice that there are no experimental results.}
\begin{tabular*}{\textwidth}{@{\extracolsep{\fill}}lcccc@{}}
\cline{2-5}
\multicolumn{1}{c}{} &   & This work & & Birkfellner~\cite{Birkfellner} \\
\cline{2-4}
\multicolumn{1}{c}{} &$\alpha_S=\beta_S=0$ & $\alpha_S=0 \text{ and }\beta_S=-1-2\alpha_S$ &  $\alpha_S=\alpha_L \text{ and }\beta_S=-1-2\alpha_S$ &\\
\hline
$K_S \to \mu \bar{\mu} \mu \bar{\mu}$ & $1.40 \times 10^{-14}$ & $1.37 \times 10^{-14}$ &$2.61\times10^{-14}$ & $1 \times 10^{-14}$ \\ 
$K_S \to e \bar{e} e \bar{e}$ & $1.66 \times 10^{-10}$ & $1.66 \times 10^{-10}$ & $1.78\times 10^{-10}$ & $7 \times 10^{-11}$ \\ 
$K_S \to \mu \bar{\mu} e \bar{e}$ &  $7.88 \times 10^{-12}$ &  $7.87 \times 10^{-12}$& $1.29\times10^{-11}$  & $8 \times 10^{-12}$   \\
\hline
\end{tabular*}\label{tab:KSdecays}
\end{center}
\renewcommand{\arraystretch}{1}
\end{table*}

\section{Interferences}

\subsection{SM CP conserving interferences}

Both determinations of the sign and of the value of $K_L \rightarrow \gamma \gamma$ are very challenging, as it has been shown in \cite{IU}. Indeed, the sign of $\mathcal{A}(K_L \rightarrow \gamma \gamma)$ is responsible for the increase or decrease of the interference contribution between short and long distance contributions in the decay $K_L \rightarrow \mu\mu$. From the CKM matrix point of view, it means that one can constrain more the $\bar{\rho}$ parameter.  We propose an experimental analysis through the interferences of $K_L$ and $K_S$ into four leptons to fix the sign.

Since the $K_L$ and $K_S$ are composite systems in the point of view of CP violation, we have to take into account this fact. It means that from now we cannot longer identify $K_1$ and $K_2$ to $K_S$ and $K_L$, but 
\begin{align}
\begin{pmatrix}
K_S \\ K_L 
\end{pmatrix}
&= \frac{1}{\sqrt{1+|\varepsilon|^2}}\begin{pmatrix}
1 & \varepsilon \\
\varepsilon & 1 
\end{pmatrix}
\begin{pmatrix}
K_1\\ K_2 
\end{pmatrix}  \\
&= \frac{1}{\sqrt{2(1+|\varepsilon|^2})}\begin{pmatrix}
1 & \varepsilon \\
\varepsilon & 1 
\end{pmatrix}
\begin{pmatrix}
1 & 1 \\
-1& 1 
\end{pmatrix}
\begin{pmatrix}
K_0\\ \bar{K}^0 
\end{pmatrix},
\end{align}
with $\text{Re }\varepsilon=1.66\times10^{-3}$ and $\text{Im }\varepsilon=1.57\times10^{-3}$. 

First, to take into account the CP asymmetry, but in the CP conserving limit ($\varepsilon=0$),  a pertinent observable to measure the oscillations between $K_S$ and $K_L$ is according to~\cite{DIPP,Heiliger:1993qt},
\begin{multline}
A^{LS} (t) \\
= \frac{2 \mathrm{e}^{- \Gamma t} \displaystyle \int \mathrm{d} \Phi_4 \, f(X,Y) \, \mathrm{Re} A_L \, \mathrm{Re} A_S}{\displaystyle \int \mathrm{d} \Phi_4 \, \left[ \mathrm{e}^{- \Gamma_S t} |A_S|^2 + \mathrm{e}^{- \Gamma_L t} |A_L|^2\right]} \cos (\Delta M \, t),
\end{multline}
for some weight function $f(X,Y)$ and we will choose here $f(X,Y) \doteq  \mathrm{sgn}(\cos \phi \sin \phi)$, 
\begin{equation}
\Delta M=  M_L - M_S \;\; \text{and} \;\; \Gamma = \frac{1}{2} \left( \Gamma_L + \Gamma_S \right).
\end{equation}

\begin{figure*}
\centering
\includegraphics[scale=0.85]{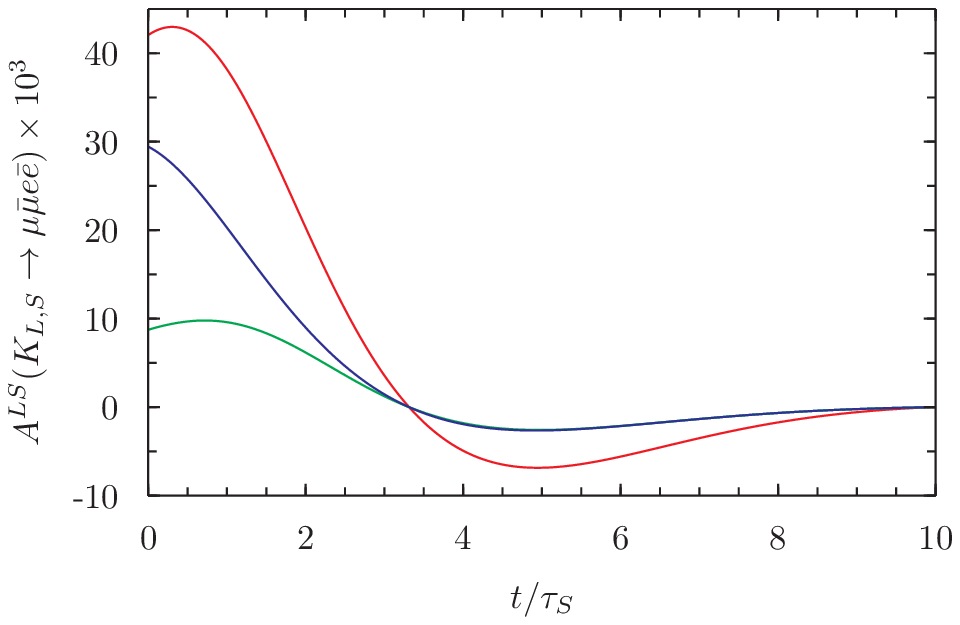} \\
\centering
\vspace*{0.8cm}
\includegraphics[scale=0.85]{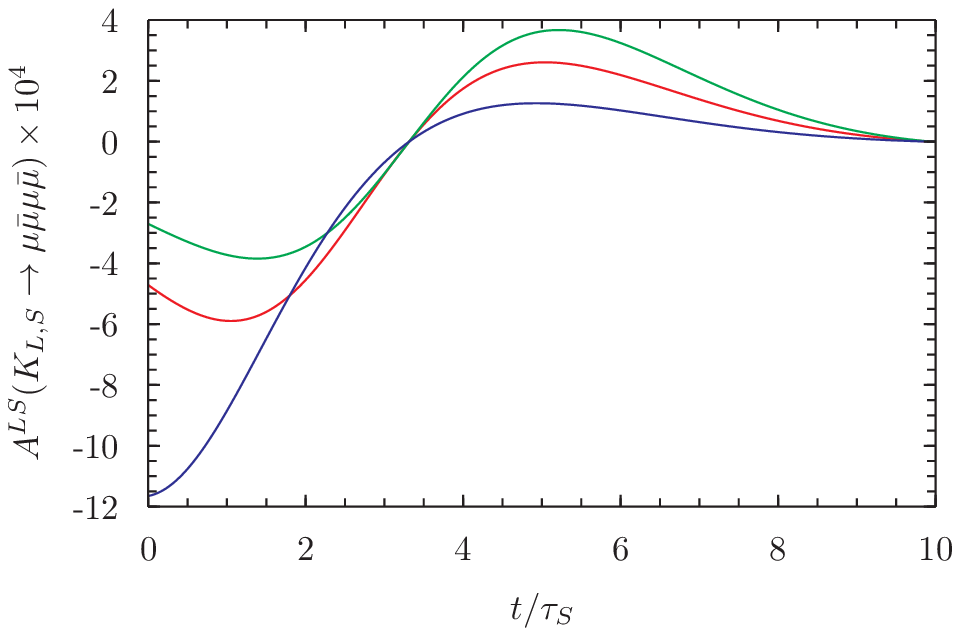}  \hspace*{0.8cm}\includegraphics[scale=0.85]{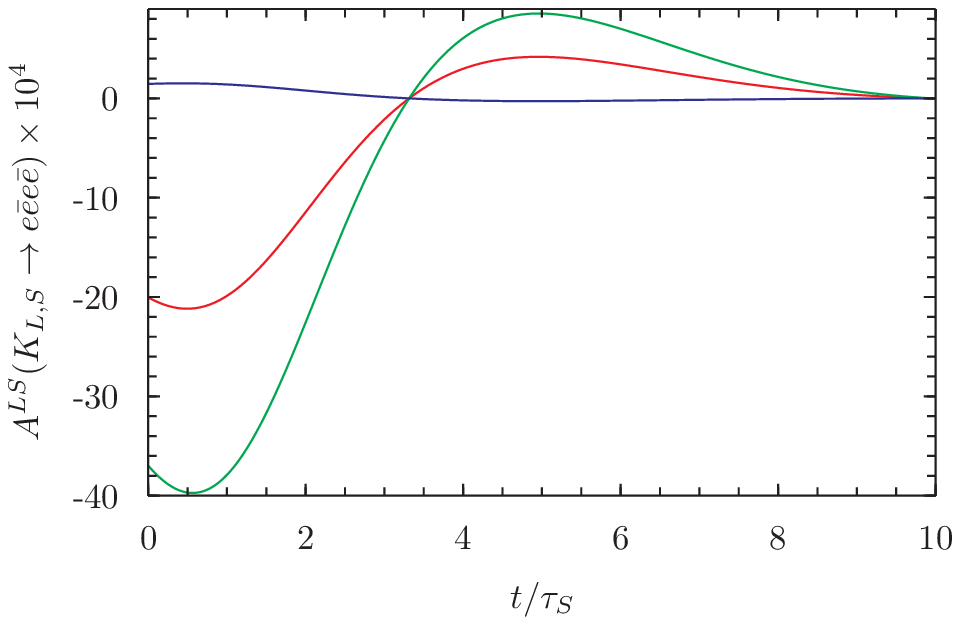}      
\caption{Interferences between $K_L$ and $K_S \to \ell_1 \bar{\ell}_1\ell_2 \bar{\ell}_2$. The red line corresponds to the case $\alpha_S=0$, the green line is $\alpha_S=-3$ while the blue line is $\alpha_S=3$. As explained in the text we assume  the sign  $K_L \rightarrow \gamma \gamma$. The interferences being directly related to this sign, their experimental observations (in shape and amplitude) could confirmed this hypothesis. }\label{fig:ALSCC}
\end{figure*}

We can easily obtain this function of time in our calculations since we can evaluate each part and we present our results for the three different channels on the fig.~\ref {fig:ALSCC}. Since $\mathrm{F}_S$ and $\mathrm{F}_L$  depend respectively  on $\alpha_S$ and $\beta_L$, we take the arbitrariness to give the plots for three different values of $\alpha_S = \{-3,0,3\}$ and using the short-distance constraint \mbox{$1+2\alpha_S+\beta_S=0$} whereas the value of $\alpha_L$ is fixed to $-1.63$ and we use the sum rule $1+2\alpha_L+\beta_L=0.3$~\cite{DAIP}.

We want to stress here that the $\alpha_L$ as the value of the slope of the form factor does not fix the sign of  $K_L \rightarrow \gamma \gamma$, as explained in the second DA$\rm \Phi$NE book  \cite{D'Ambrosio:1994ae}, it keeps an ambiguity. If we assume the VMD model for the weak form factor, this ambiguity can be removed as it has been shown in \cite{DAmbPorto} and confirmed by other theoretical considerations in \cite{Gerard:2005yk}.  In our approach, we take that the dominant low energy contribution is coming from the pion pole, implying then
\begin{equation}
\text{sgn} [\mathcal{A}(K_L \rightarrow \gamma \gamma )] =\text{sgn} [\mathcal{A}(K_L \rightarrow \pi^0 \rightarrow \gamma \gamma )]\;,
\end{equation}    
thus  we will do all the following analysis under this statement. But of course, the experimental interferences analysis that we propose allows us to remove the ambiguity since the shape is fixed by the sign of $K_L \rightarrow \gamma \gamma$.

\subsection{NP contributions to CP violation interferences}

As matter of principles, one can question an eventual apparition of New Physics contributions to these decays. Of course, to be fully descriptive we have first taken into account all already permitting contributions and evaluated their size to pretend to see new signatures in experimental results. This is the reason why we decompose all the possible contributions to the amplitude of the decay of $K_S$ as\footnote{We are aware that we do a misuse of writing by exponentiating the amplitude since we do not prove any unitarization of the amplitudes as long as  we consider only the first term. But it is quite obvious that faced with the smallness of the numbers, this cannot change a lot the conclusions.} 
\begin{align}
A_S = |A_1|  \mathrm{e}^{i \delta} + \varepsilon  |A_2| \mathrm{e}^{i \delta'} +
 i |A_1^B| + i |A_1'|  \mathrm{e}^{i \delta'},
\end{align}
viz.
\begin{itemize}
\item $|A_1|  \mathrm{e}^{i \delta} $ is the $K_S$ CP conserving part. $A_1$ is just the SM amplitude computed in the section~\ref{KS}.  $\delta$ is related through the optical theorem to the absorptive part of $\mathcal{A}(K_S \to \pi \pi \to \gamma^* \gamma^*)$. It is given by \cite{Ecker:1991ru}:
\begin{equation}
\arctan \delta =  \frac{\mathrm{Im} \, \mathcal{A}(K_S \to \pi \pi\to \gamma^* \gamma^*)}{\mathrm{Re} \, \mathcal{A}(K_S \to \pi \pi\to \gamma^* \gamma^*)},
\end{equation} 
with:
\begin{align}
&\mathrm{Re} \, \mathcal{A}(K_S \to \pi \pi\to \gamma^* \gamma^*) \nonumber\\
&\hspace*{0.5cm}= -\frac{1}{2} \left\{ 1 + r_\pi^2 \left[ \ln^2 \left( \frac{1-\sigma_\pi}{1+\sigma_\pi} \right)+ \pi^2 \right] \right\},
\end{align}

\begin{multline}
\mathrm{Im} \, \mathcal{A}(K_S \to \pi \pi\to \gamma^* \gamma^*)  \\
=  - \pi \, r_\pi^2 \ln \left( \frac{1-\sigma_\pi}{1+\sigma_\pi} \right),
\end{multline}
where $r_\pi = M_\pi/M_K$ and $\sigma_\pi=\sqrt{1-4 r_\pi^2}$.
This yields to $\delta \approx -27.54^{\circ}  \approx -0.48065$.\\

\item $\varepsilon  |A_2| \mathrm{e}^{i \delta'} $ is the $K_L $ CP-violating part. $A_2$ is the SM amplitude and  $\varepsilon$ parametrizes the indirect CP violation. $\delta'$ is related through the optical theorem to the absorptive part of $\mathcal{A}(K_L \to \mu \bar{\mu})$. It is given by:
\begin{equation}
\arctan \delta' =  \frac{\mathrm{Im} \, \mathcal{A}(K_L \to \mu \bar{\mu})}{\mathrm{Re} \, \mathcal{A}(K_L \to \mu \bar{\mu})},
\end{equation}
with~\cite{GDP,Knecht:1999gb}:

\begin{align}
&\mathrm{Re} \, \mathcal{A}(K_L \to \mu \bar{\mu}) \nonumber\\  
&\hspace*{0.5cm}= \frac{1}{4 \sigma_\mu} \ln^2\left( \frac{1-\sigma_\mu}{1+\sigma_\mu}\right) + \frac{1}{\sigma_\mu} \mathrm{Li}_2 \left( \frac{\sigma_\mu-1}{\sigma_\mu+1}\right) \nonumber\\
&\hspace*{0.6cm}+ \frac{\pi^2}{12 \sigma_\mu} + 3 \ln \frac{m_\mu}{\mu} + \chi(\mu),
\end{align}

\begin{equation}
\mathrm{Im} \, \mathcal{A}(K_L \to \mu \bar{\mu}) =  \frac{\pi}{2 \sigma_\mu} \ln \left( \frac{1-\sigma_\mu}{1+\sigma_\mu}\right),
\end{equation}
where $r_\mu = m_\mu/M_K$, $\sigma_\mu=\sqrt{1-4 r_\mu^2}$ and\\ $\chi(M_\rho)=3.3$.
We get  $\delta' \approx -82.48^{\circ} \approx -1.43952$.\\

\item $i|A_1^B|$ is the $K_S$ Bremsstrahlung CP-violating part. For more details see the appendices.\\

\item $|A_1'|  \mathrm{e}^{i \delta'}$ is the $K_S$  CP-violating part. $A_1'$ is a CP-violation part of the $K_S$ coming from a coupling to the photons similar to the one of the $K_L$: \\$K_S \, \varepsilon_{\mu \nu \rho \sigma} F^{\mu \nu} F^{\rho \sigma}$. Consequently $A_1'$ would be proportional to $A_2$. An estimation of the relation between  $A_1'$ and $A_2$ gives: $A_1' \doteq \xi A_2$ with $\xi \sim10^{-1}$ \cite{IU}. The strong phase $\delta'$ is the same as for the second term due to universality, the coupling of the $K_{S,L}$ to the two photons in this case being similar. This constitutes our NP implementation. 
\end{itemize}

Contrary to usual asymmetries prescriptions to have a relevant observable to distinguish the most important contribution, in the case of identical leptons pairs, we have to consider the following phase space integration
\begin{equation}
\int_0^\pi \mathrm{d} \phi\int \, \frac{\mathrm{d} \Phi_4 }{\mathrm{d}\phi }\, \mathrm{sgn}(\cos \phi \sin \phi) |A_S|^2.
\end{equation}

\begin{figure}
\includegraphics[scale=0.8]{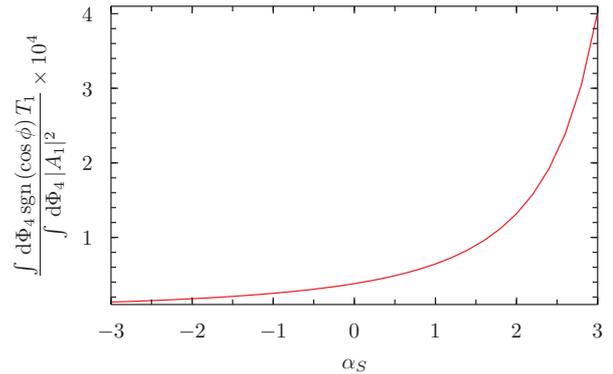}
\caption{Dominant NP contribution to $K_S \rightarrow 4\mu$ for the direct CP violation contribution and where  we use the short notation $T_1= \left( |\varepsilon|^2 + \xi^2 + 2 \xi\, \rm{Im\,} \varepsilon \right)|A_2|^2  $.}\label{fig:T1term}
\end{figure}

A straightforward computation of all parts, illustrated on the fig.~\ref{fig:T1term} only for the channel into four muons, show the $A'_1$ NP part is dominant as expected,  assuming that it is universal and just an approximation to the dominant behaviour.

\begin{figure*}
\centering
\includegraphics[scale=0.85]{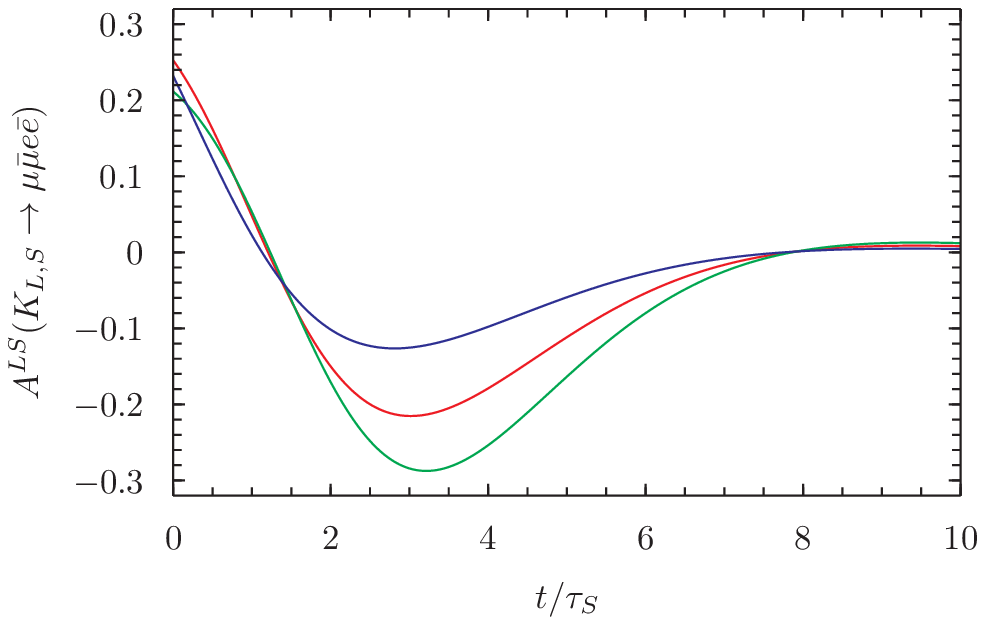}  \\ 
\centering
\vspace*{0.8cm}
\includegraphics[scale=0.85]{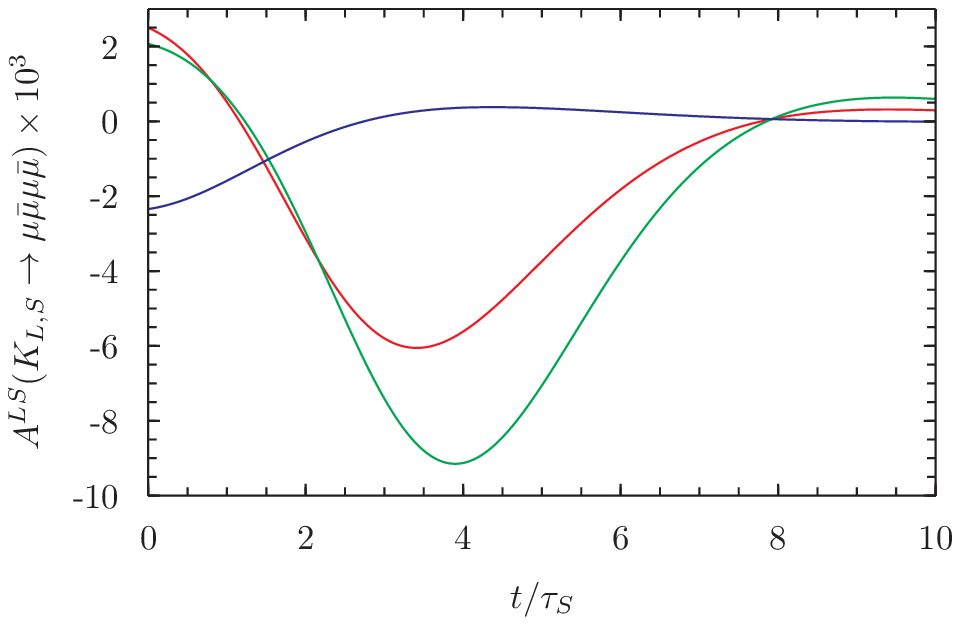} 
\hspace*{0.5cm}
\includegraphics[scale=0.85]{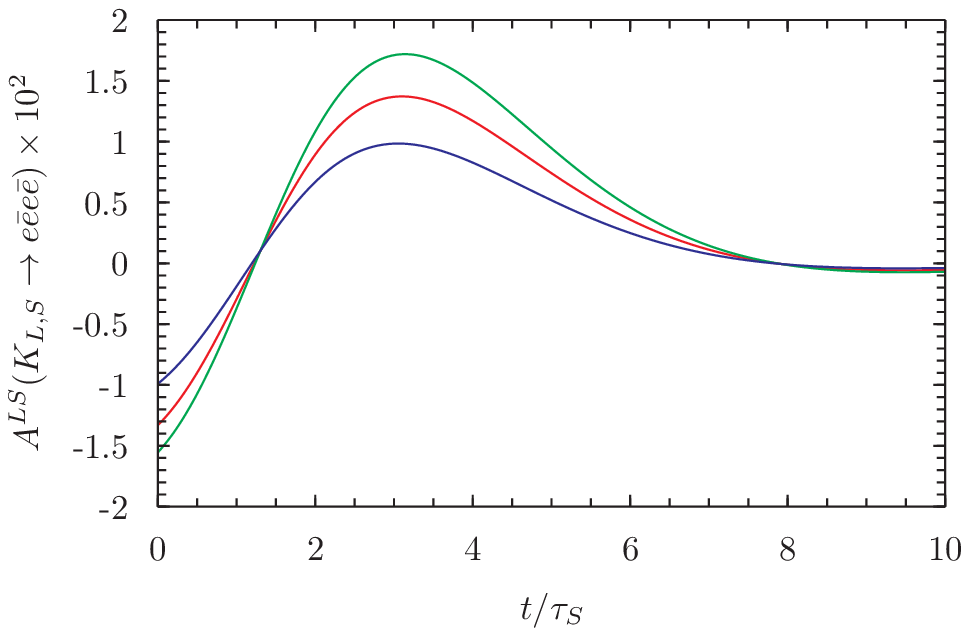} 
\caption{New Physics contributions to the CP violation interferences. The red line corresponds to the case $\alpha_S=0$, the green line is $\alpha_S=-3$ while the blue line is $\alpha_S=3$. . Here too we make the same assumption for the sign of $K_L \rightarrow \gamma \gamma$.}\label{fig:ALSNP}
\end{figure*}

Since the hypothesis of the dominant part is coming from the NP contribution one can suppose  now to  generate as an observable $A^{LS} (t)$ such as

\begin{strip}
\begin{equation}
A^{LS} (t) = \frac{\mathrm{e}^{- \Gamma t} \displaystyle \int_0^{\phi_0} \mathrm{d} \phi\int \, \frac{\mathrm{d} \Phi_4 }{\mathrm{d}\phi }\, \mathrm{sgn}(\cos \phi \sin \phi)  \, \bigg[ \mathrm{Re} (A_L  A_S^*) \cos (\Delta M \, t) \\
+  \mathrm{Im} (A_L  A_S^*) \sin (\Delta M \, t) \bigg] } {\displaystyle \int_0^{\phi_0}  \mathrm{d} \phi\int \, \frac{\mathrm{d} \Phi_4 }{\mathrm{d}\phi }\, \left[ \mathrm{e}^{- \Gamma_S t} |A_S|^2 + \mathrm{e}^{- \Gamma_L t} |A_L|^2\right]},
\end{equation}
\end{strip}
where
\begin{eqnarray*}
A_L & = &  |A_2|, \\
A_S & = &  |A_1| \mathrm{e}^{i \delta} + i |A_1'|  \mathrm{e}^{i \delta'} =  |A_1| \mathrm{e}^{i \delta} + i \xi  |A_2| \mathrm{e}^{i \delta'}
\end{eqnarray*}
and $\phi_0$ is the angle that maximizes $A^{LS} (t)$, $\phi_0=\pi$ for the case where the four leptons are identical and  $\phi_0=\pi/2$ for $K_S \rightarrow \bar \mu \mu \bar e e$~\cite{Sehgal:1999vg,CCDA}.

Therefore one obtains the  results of fig.~\ref{fig:ALSNP}.  It is obvious that they results expected in our calculation are made under the assumptions of the sign of the amplitude $\mathcal{A}(K_L\rightarrow \gamma \gamma)$, if experimentally one observes the same kind of curves (after fixing $\alpha_L$ from the decays) it means that our hypothesis for the sign is correct, if the curves are symmetric about the horizontal axis it implies the opposite sign naturally.

\section{Conclusions}

We have shown that it is possible to obtain good predictions for the branching ratios for the decays of the $K_L$ into four leptons comparing to the experimental data through a  vector meson dominance inspired form factor. It is natural then to consider that the same approach is pertinent for the case of the $K_S$ into four leptons, since the model is more constrained from short distance behaviour. Since this short distance behaviour is  model dependant in our approach, one can emphasize that even if our assumptions of a VMD form factor type, one could ever  consider the slope ($\alpha$) by itself and see it as the first derivative of the form factor experimentally observe in a model independent way. 

A direct consequence of the experimental data in our approach would be to fix the $\alpha$ and $\beta$ parameters for $K_L$ and $K_S$ and then give  the sign of  $\mathcal{A}(K_L\to \gamma\gamma)$ (for a sufficient accuracy of course). Moreover, we have shown that a simple assumption on the existence of  a NP operator in the lagrangian could be verified with interferences in those decays. 

It appears that now it is important to obtain some experimental data in these channels involving the $K_S$ decays (particularly the muons ones)  and considering our predictions, we hope that the LHCb processes for tagging the muons allow us to reach a sufficient level of accuracy. We think also, that it could be easier to identify the decays rates containing electrons through the ones involving pions decays.

\begin{acknowledgements}
The authors would like to thank  F. Ambrosino, O. Cata, P. Massarotti, J. Portol\'es for his careful reading of the manuscript, E. de Rafael and M. D. Sokoloff. G.D.  acknowledges partial support by MIUR under project 2010YJ2NYW \\ (SIMAMI). D.G.'s work is supported in part by the EU under Contract MTRN-CT-2006-035482 (FLAVIAnet) and by MUIR, Italy, under Project 2005-023102. G.V.'s research  has been supported in part by the Spanish Government and ERDF funds from the EU Commission [grants FPA2007-60323, CSD2007-00042 (Consolider Project CPAN)].
\end{acknowledgements}

\appendix 

\section{Detailed expressions for amplitudes}

\begin{figure}[h!]
\centering
\includegraphics[scale=0.575]{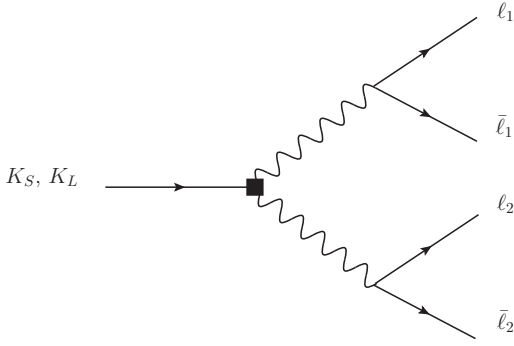}
\caption{Amplitudes of $K_{S,L}$ in four leptons.}
\end{figure}

\subsection{The $K_S$ decays amplitudes}

For the cases where $\ell_1=\ell_2$, there exist 4 diagrams that can be reduced to two different amplitudes $\mathcal{M}_A$ and $\mathcal{M}_B$, 
\begin{align}
\mathcal{M}_A & = e^2 \frac{\mathrm{F}_S\left((p_1+p_2)^2,(p_3+p_4)^2\right)}{(p_1+p_2)^2 (p_3+p_4)^2}  \nonumber \\
&\times   \left[ (p_1+p_2) \cdot (p_3+p_4) g^{\mu \nu} -  (p_1+p_2)^\nu (p_3+p_4)^\mu \right] \nonumber\\
&\times\left[ \bar{u}(p_1) \gamma_\mu  v(p_2) \right]  \left[ \bar{u}(p_3)  \gamma_\nu  v(p_4) \right]
\end{align}
and
\begin{align}
\mathcal{M}_B & = -e^2 \frac{\mathrm{F}_S\left((p_3+p_2)^2,(p_1+p_4)^2\right)}{(p_3+p_2)^2 (p_1+p_4)^2}  \nonumber \\
&\times   \left[ (p_2+p_3) \cdot (p_1+p_4) g^{\mu \nu} -  (p_2+p_3)^\nu (p_1+p_4)^\mu \right] \nonumber \\
& \times\left[ \bar{u}(p_3) \gamma_\mu  v(p_2) \right] \left[ \bar{u}(p_1)  \gamma_\nu  v(p_4) \right].
\end{align}
Thus the total squared amplitude is given by (under symmetries considerations, $|\mathcal{M}_A|^2 = |\mathcal{M}_B|^2 $),
\begin{equation}
| \mathcal{M} |^2 = \frac{1}{2}(|\mathcal{M}_A|^2  + \mathcal{M}_A \, \mathcal{M}_B^*).
\end{equation}

In the mixed case, $\ell_1=\mu$ and $\ell_2=e$, there are only two diagrams, and we have 
$|\mathcal{M}|^2=|\mathcal{M}_A|^2$.

\subsection{The $K_L$ decays amplitudes}

The calculations of the amplitudes involving the $K_L$ are identical in procedure that the ones for the $K_S$, we have to distinguish two kinds of amplitudes 
\begin{align}
\mathcal{M}_A & = e^2 \frac{\mathrm{F}_L\left((p_1+p_2)^2,(p_3+p_4)^2\right)}{(p_1+p_2)^2 (p_3+p_4)^2}  \nonumber \\
&\times    \varepsilon_{\mu \nu \rho \sigma } (p_1+p_2)^\rho (p_3+p_4)^\sigma  \nonumber\\
&\times\left[ \bar{u}(p_1)  \gamma^\mu  v(p_2) \right]  \left[ \bar{u}(p_3) \gamma^\nu  v(p_4) \right]
\end{align}
and
\begin{align}
\mathcal{M}_B & =- e^2 \frac{\mathrm{F}_L\left((p_3+p_2)^2,(p_1+p_4)^2\right)}{(p_3+p_2)^2 (p_1+p_4)^2}  \nonumber \\
&\times   \varepsilon_{\mu \nu \rho \sigma } (p_2+p_3)^\rho (p_1+p_4)^\sigma\nonumber \\
&\times  \left[ \bar{u}(p_3) \gamma^\mu v(p_2) \right] \left[ \bar{u}(p_1) \gamma^\nu  v(p_4) \right].
\end{align}
The total amplitude is given by (under symmetries considerations, $|\mathcal{M}_A|^2 = |\mathcal{M}_B|^2 $),
\begin{equation}
| \mathcal{M} |^2 = \frac{1}{2}(|\mathcal{M}_A|^2  + \mathcal{M}_A \, \mathcal{M}_B^*).
\end{equation}
As before, in the mixed case, $\ell_1=\mu$ and $\ell_2=e$, there are only two diagrams, and we have $|\mathcal{M}|^2=|\mathcal{M}_A|^2$.

\section{Bremsstrahlung CP-violating part}

\begin{figure}[h!]
\centering
\includegraphics[scale=0.575]{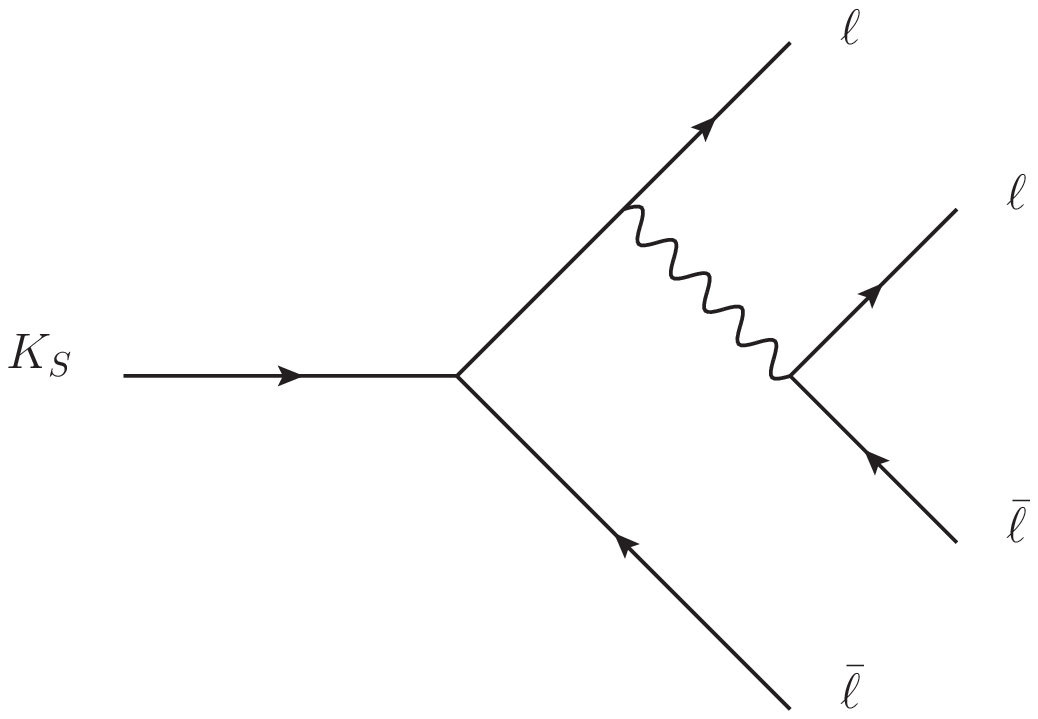}
\includegraphics[scale=0.575]{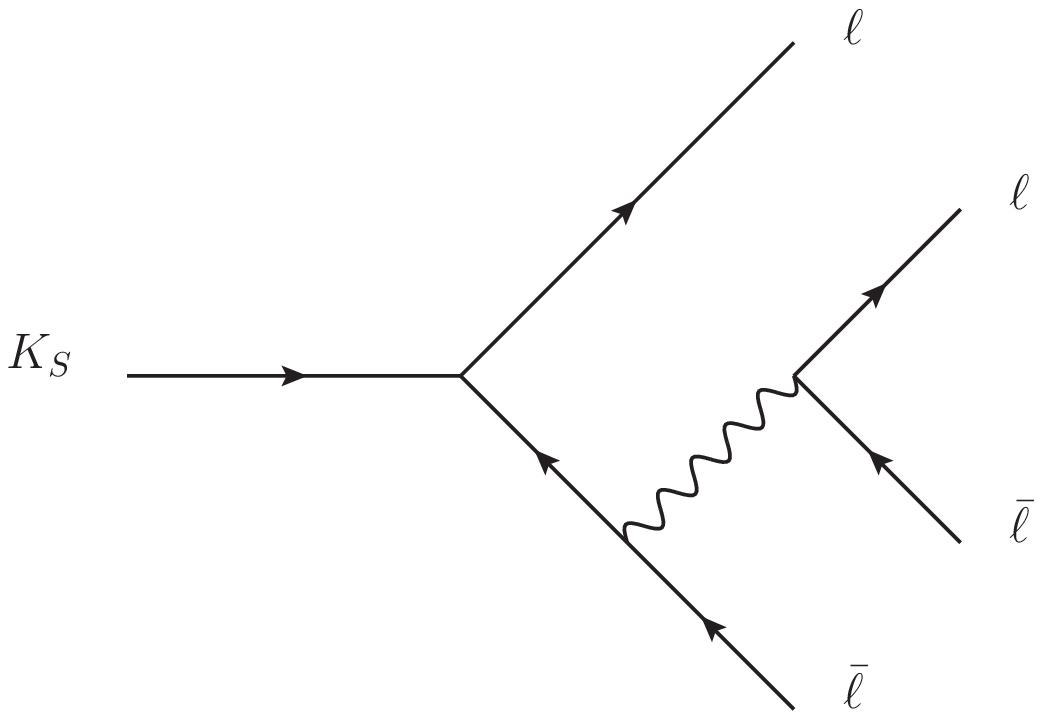}
\caption{Bremsstrahlung amplitudes for $K_S$ in four leptons.}
\end{figure}

Using the Low's theorem~\cite{Low:1958sn}, the amplitude\\
$K_S \, (q)\to \mu \, (p_-)  \, \bar{\mu} \, (p_+) \, \gamma^* \,(k)$ is the product of the\\
 $K_S \, (q)\to \mu \, (p_-)  \, \bar{\mu} \,(p_+)$ amplitude times the contribution of the soft photon radiated:
\begin{multline}
\mathcal{M}(K_S \to \mu \bar{\mu} \gamma^*)  \\
=k^2\mathrm{F}_B^\mu (k,p_-,p_+)\;\epsilon_{\mu}(k) \mathcal{M}(K_S \to \mu \bar{\mu}),
\end{multline}
where $ \mathcal{M}(K_S \to \mu \bar{\mu})$ is the decay amplitude of $K_S$ into two muons
\begin{equation}
 \mathcal{M}(K_S \to \mu \bar{\mu}) = \mathcal{A}_{SD} \; \bar{u} \gamma_5 v
\end{equation}
 and 
\begin{multline}
\mathrm{F}_B^\mu (k,p_-,p_+) \\
= \frac{2e}{k^2} \left[ \frac{p_-^\mu}{k^2 +2 k \cdot p_-} + \frac{p_+^\mu}{k^2 -2 k \cdot p_+}\right].
\end{multline}

Now, we just have to contract with the muonic current $-i e \bar{u}\gamma^\nu v\;\epsilon^*_\nu$ to obtain the Bremsstrahlung contribution
\begin{align}
&\mathcal{M}_\text{Brems.} \nonumber\\
&= \mathcal{A}_{SD} \bigg \{  \mathrm{F}_B^\mu (q_1,p_3,p_4) \left[ \bar{u}(p_1) \gamma_\mu v(p_2) \right] \, \left[ \bar{u}(p_3) \gamma_5 v(p_4) \right]  \nonumber\\
&  + \mathrm{F}_B^\mu (q_2,p_1,p_2) \left[ \bar{u}(p_3) \gamma_\mu v(p_4) \right] \, \left[ \bar{u}(p_1) \gamma_5 v(p_2) \right] \nonumber \\
 & - \mathrm{F}_B^\mu (p_2+p_3,p_1,p_4) \left[ \bar{u}(p_3) \gamma_5 v(p_2) \right]  \left[ \bar{u}(p_1) \gamma_\mu v(p_4) \right]  \nonumber\\
 &  - \mathrm{F}_B^\mu (p_1+p_4,p_3,p_2) \left[ \bar{u}(p_1) \gamma_\mu v(p_4) \right]\left[ \bar{u}(p_3) \gamma_5 v(p_2) \right] \Bigg\}.
\end{align}

In our case, $\mathrm{Re} \, \mathcal{A}_{SD} $ can be neglected, all the short-distance information is contained in  $\mathrm{Im} \, \mathcal{A}_{SD} $ and we have~\cite{IU}:
\begin{multline}
\mathrm{Im} \,\mathcal{A}_{SD}  \\
= - \frac{G_F \, \alpha_\mathrm{em}(M_Z)}{\pi \sin^2 \theta_W}\sqrt{2} m_\mu F_K \mathrm{Im} \, (V_{ts}^* V_{td}) Y(x_t),
\end{multline}
with $x_t=m_t^2/M_W^2$ and $Y(x)$ is the Inami-Lin function:
\begin{equation}
 Y(x) = \frac{x}{8}\left[\frac{4-x}{1-x} + \frac{3x}{(1-x)^2} \ln x\right].
\end{equation}


\end{document}